\documentclass[amsmath,amssymb,superscriptaddress,prd,nofootinbib,twocolumn]{revtex4-2}
\usepackage{graphicx}
\usepackage{amsmath}
\usepackage{amssymb}
\usepackage{amsfonts}
\usepackage{color}
\usepackage{dsfont}
\usepackage{tikz}
\usepackage{ragged2e}
\usetikzlibrary{calc,decorations.markings}
\usepackage[colorlinks=true,linkcolor=blue,citecolor=blue,urlcolor=blue]{hyperref}
\usepackage{physics}
\usepackage{soul}
\usepackage{float}
\usepackage{subfig}
\newcommand{\xx}{\mathbf{x}}
\newcommand{\x}{\mathsf{x}}
\usepackage{tikz,xcolor}
\definecolor{lime}{HTML}{A6CE39}
\DeclareRobustCommand{\orcidicon}{%
	\begin{tikzpicture}
	\draw[lime, fill=lime] (0,0) 
	circle [radius=0.16] 
	node[white] {{\fontfamily{qag}\selectfont \tiny ID}};
	\draw[white, fill=white] (-0.0625,0.095) 
	circle [radius=0.007];
	\end{tikzpicture}
	\hspace{-2mm}
}
\foreach \x in {A, ..., Z}{%
	\expandafter\xdef\csname orcid\x\endcsname{\noexpand\href{https://orcid.org/\csname orcidauthor\x\endcsname}{\noexpand\orcidicon}}
}

\begin{document}
\title{Quantum thermal machines in BTZ black hole spacetime}
\author{Dimitris Moustos\orcidA{}}
\email{dimitris.moustos@newcastle.ac.uk}
\affiliation{School of Mathematics, Statistics, and Physics, Newcastle University, Newcastle upon Tyne NE1 7RU, United Kingdom}
\author{Obinna Abah\orcidB{}}
\email{obinna.abah@newcastle.ac.uk}
\affiliation{School of Mathematics, Statistics, and Physics, Newcastle University, Newcastle upon Tyne NE1 7RU, United Kingdom}
\date{\today}

\begin{abstract}
We investigate an Otto thermodynamic cycle with a qubit Unruh-DeWitt  detector as the working medium, coupled to a massless, conformally coupled scalar quantum field in the Hartle-Hawking vacuum in a (2+1)-dimensional BTZ black hole spacetime. We employ the thermal properties of the field to model heat and cold reservoirs between which the thermal machine operates. Treating the detector as an open quantum system, we employ a master equation to study its finite-time dynamics during each cycle stroke. We evaluate the output performance of the Otto heat engine and refrigerator by computing, respectively, the total work output and the cooling power for each of the Neumann, transparent, and Dirichlet boundary condition cases satisfied by the field at spatial infinity. Furthermore, we evaluate the optimal performance of the thermal machine by analyzing its efficiency at maximum power output and ecological impact. Our study presents a general framework for understanding the finite-time operation of relativistic quantum thermal machines, focusing on their energy optimization. 
\end{abstract}

\maketitle

\section{Introduction}
The technological advances achieved in the
past decade in the fabrication of miniaturized nanoscale devices, along
with the rapid development of quantum thermodynamics \cite{vinjanampathy2016quantum,binder2019thermodynamics,Deffner:Campell,potts2024quantumthermodynamics}, have paved the way for the design of novel thermal machines that operate in the quantum regime \cite{therm:eng:rev}. Among them, the quantum Otto cycle \cite{Abah_2012,Abah_2016,Kosloff} -- comprising two adiabatic and two isochoric processes -- is the most widely studied because it allows an explicit identification of heat and work exchanges in each process step. When operating as a heat engine, the Otto machine converts absorbed heat from a hot reservoir to work, while as a refrigerator, it consumes work to extract heat from a cold reservoir.

The second law of thermodynamics imposes a universal upper bound on the efficiency of heat engines, given by the Carnot efficiency \cite{Callen}, $\eta_C=(T_h-T_c)/T_h$, where $T_h$ and $T_c$ are the temperatures of the hot and cold reservoirs respectively. Similarly, the maximum coefficient of performance (COP) for refrigerators is $\varepsilon_C=T_c/(T_h-T_c)$ \cite{Callen}. However, these Carnot bounds are only obtained in the
limit of infinitely slow strokes. More relevant for real thermal machines that operate in finite time are the efficiency at maximum power (EMP) \cite{Curzon:Ahlborn,Deffner:EMP}, $\eta^*$, for heat engines, and the COP at maximum figure of merit \cite{Abah_2016}, $\varepsilon^*$, for refrigerators.

In quantum field theory, the notions of a vacuum state and particle number become observer-dependent in curved spacetimes or for non-inertial observers \cite{Birrell:Davies}. This is exemplified by the Hawking \cite{Hawking} and Unruh \cite{Unruh} effects. The Unruh effect states that a uniformly  accelerating observer perceives the Minkowski vacuum state of a quantum field as a thermal state, with a temperature proportional to their acceleration. Similarly, the Hawking effect describes how an observer near a black hole horizon detects a thermal spectrum of particles—the Hawking radiation—emitted by the black hole due to quantum vacuum fluctuations. Hawking and Unruh effects reveal a close connection between quantum physics, relativity, and thermodynamics.

An operational way to probing the particle content of a quantum field is through the Unruh-DeWitt (UDW) detector \cite{Unruh,DeWitt,Hu_Louko}, a  pointlike  two-level quantum system (qubit) that locally interacts with the field. Excitations of the detector are interpreted to result from the absorption of field quanta, i.e., particles. Initially introduced to probe the particle phenomenology in the Unruh and Hawking effects, nowadays the UDW detector model is widely employed in relativistic quantum information \cite{Mann_2012}, having found many applications in implementing quantum information protocols in relativistic settings.

The thermal response of uniformly accelerated detectors has been recently used to introduce the concept of an Unruh quantum Otto heat engine \cite{arias2018unruh,gray2018,XU2020135201,Unruh:entangl,barman2022,Mukherjee_2022}, where the  detector serves as the working medium, undergoing an Otto thermodynamic cycle. During the isochoric stages of the cycle, the detector maintains constant acceleration while  interacting weakly with a quantum field, which, due to the Unruh effect, effectively behaves as a thermal reservoir. Thus, by adjusting the magnitude of acceleration, one can simulate hot and cold heat reservoirs, allowing the engine to extract useful work. 

Later studies have expanded upon the concept of the Unruh Otto heat engine by introducing a more general model of a relativistic quantum Otto engine, which builds on
the effective temperatures observed by detectors following stationary trajectories while interacting with quantum fields in Gaussian states in curved spacetime backgrounds \cite{gallock2023quantum}. In this setup, the work that can be extracted by a detector following a circular trajectory has been specifically investigated. Moreover, additional studies explore the effects of the relativistic motion of detectors through thermal field baths \cite{papadatos2021quantum,NK:DM}, as well as 
instantaneous detector-field interactions \cite{NK:DM,gallock2024relativistic}, on the performance of the Otto engine and the amount of extracted work.

In this paper, we utilize the thermal response of a UDW detector to a quantum field in a black hole background, to model the field as a heat reservoir and introduce a quantum Otto thermodynamic cycle. In particular, during the isochoric strokes of the Otto cycle, the detector interacts with a massless, conformally coupled scalar quantum field in the Hartle-Hawking vacuum state in a (2+1)-dimensional Bañados-Teitelboim-Zanelli (BTZ) black hole spacetime \cite{BTZ1,BTZ2}. The Wightman two-point corelation function of the conformally coupled field in this case is analytically evaluated \cite{Lifschytz:Ortiz,Carlip:1995}, facilitating the investigation of the detector dynamics. This has been employed, for example, in studying transition rates \cite{Hodg2012,PreciadoRivas:2024,spadafora2024deepknottedblackhole}, the response of a detector to a black hole in a superposition of masses \cite{Foo}, and the process of extracting correlations from the quantum field vacuum \cite{henderson2018harvesting, Gallock:Mann:Harv}. Related studies have investigated the performance of quantum batteries--modeled as UDW detectors--in the BTZ spacetime \cite{Tian:2024wby}, as well as the efficiency of a Brayton thermodynamic cycle using the BTZ black hole as a working medium \cite{Ferketic:Deffner}. Besides, a mechanism functioning as a black hole-powered quantum heat engine has been proposed in \cite{misra2024black}, where redirected Hawking radiation excites atoms falling into a black hole, enabling coherent light amplification and effectively converting vacuum energy into work. 

We treat the UDW detector as an open quantum system, with the conformally coupled field playing the role of the environment, and describe its time evolution using a master equation of the Gorini–
Kossakowski–Sudarshan–Lindblad (GKSL) form \cite{breuer}. By employing a finite-time analysis of the detector's dynamics, we explicitly investigate how the thermal machine's performance depends on the Neumann, transparent, and Dirichlet boundary conditions satisfied by the field at spatial infinity. This analysis is not possible in the long-time asymptotic limit, where the detector's reduced density matrix reaches a unique thermal equilibrium state, leading to identical behavior for all three boundary conditions. 

In what follows, we consider the Otto thermal machine to operate either as a heat engine or as a refrigerator. In the first scenario, we evaluate the total work output of the engine,  optimizing its performance by analyzing its EMP. When the machine functions as a refrigerator, we assess its cooling power and optimize its performance by studying the COP at the maximum figure of merit \cite{Abah_2016,Kosloff}. Our analysis identifies the Neumann boundary condition as the optimal choice for finite-time operations of the machine. In the heat engine case, it yields the highest work output among the three boundary conditions, with an efficiency at maximum power that exceeds the Curzon-Ahlborn efficiency \cite{Curzon:Ahlborn}. For the refrigerator case, it enables the absorption of the maximum amount of heat from the cold reservoir. Additionally, the Neumann boundary condition allows the detector to achieve thermalization with the hot and cold reservoirs at earlier times than the other two cases. In addition, we study the optimal performance of the Otto cycle in terms of the ecological criterion \cite{Angulo:Brown}, which balances the trade-off between high power output and low entropy production.

The paper is organized as follows. In Sec. \ref{sec:UDW:model}, we introduce the UDW particle detector model and then present its time evolution in the framework of open systems. In Sec. \ref{sec:btz}, we investigate the dynamics of a detector interacting with a conformally coupled scalar field in the BTZ black hole spacetime, analyzing how it is affected by the different boundary conditions imposed on the quantized field. In Sec \ref{BTZ:Otto}, we introduce a quantum Otto thermal machine in the BTZ spacetime, where the detector serves as the working medium, and evaluate its overall performance  when operating as either a heat engine or a refrigerator. Finally, in Sec. \ref{conclusions}, we summarize and discuss our results.
Throughout the paper we denote spatial vectors with boldface letters $(\xx)$, while spacetime vectors are represented by sans-serif characters $(\x)$. Unless otherwise specified we set $\hbar=c=k_B=1$.

\section{The UDW detector model}\label{sec:UDW:model}
We consider an Unruh-DeWitt detector \cite{Unruh,DeWitt,Hu_Louko}, modeled as a pointlike two-level quantum system (qubit) having a ground state $\ket{g}$ and an excited state $\ket{e}$, separated by an energy gap $\Omega$. The Hamiltonian that describes the qubit detector is
\begin{equation}
    H_0=\frac{\Omega}{2}\sigma_3,
\end{equation}
where $\sigma_3$ is the Pauli-Z matrix. The detector moves along a worldline $\mathsf{x}(\tau)$ parametrized by its proper time $\tau$, and is linearly coupled to a quantum field $\Phi(\x)$. In the interaction picture, the Hamiltonian that describes the interaction between the detector and the field reads
\begin{align}\label{UDW:Hamilt}
H_{\text{int}}(\tau)=\lambda \mu(\tau)\Phi(\mathsf{x}(\tau)),
\end{align}
where $\lambda$ is a small coupling constant, 
\begin{align}
\mu(\tau)=e^{i\Omega\tau}\sigma_++e^{-i\Omega\tau}\sigma_-,
\end{align}
is the detector's monopole moment operator expressed in
terms of the ladder operators $\sigma_+=\ket{e}\bra{g}$ and $\sigma_-=\ket{g}\bra{e}$ of the Pauli algebra, and $\Phi(\mathsf{x}(\tau))$ is the field evaluated along the detector’s worldline.
The interaction Hamiltonian, Eq. \eqref{UDW:Hamilt}, represents a special case of the spin-boson model Hamiltonian \cite{breuer}. This suggests the treatment of a UDW detector as an open quantum system, with the quantum
field playing the role of the environment inducing dissipation, noise, and decoherence (see, e.g., \cite{benatti,DMCA,BJA:DM,Yu:Zhang,Kaplanek,Kaplanek:Tjoa}).

\subsection{Open system dynamics}\label{sec:open:dyn}
Now, we consider that the quantum field is prepared in a state $\ket{\Psi}$, and introduce the Wightman two-point correlation function of the field,
\begin{align}\label{Wightman:fun}
    \mathcal{W}(\tau,\tau'):=\expval{\Psi|\Phi(\x(\tau))\Phi(\x(\tau'))|\Psi},
\end{align} 
evaluated along the detector's worldline. The correlation function, Eq. \eqref{Wightman:fun}, incorporates all the information related to the state of the field, the trajectory of the detector and the structure of the background spacetime.
We next focus on detectors that follow stationary trajectories and fields that are prepared in stationary states. In these cases, the Wightman function, Eq. \eqref{Wightman:fun}, depends only on the proper time deference $\tau-\tau'$ between two points on the detectors' worldline, i.e., $\mathcal{W}(\tau,\tau')=\mathcal{W}(\tau-\tau')$ \cite{Birrell:Davies,Letaw}. 

We introduce the Laplace transform, at imaginary argument $i\Omega$, of stationary Wightman functions,
\begin{align}
    \widehat{\mathcal{W}}(\Omega)\equiv\int_0^{\infty}ds e^{-i\Omega s}\mathcal{W}(s)=\frac{\Gamma(\Omega)}{2}+iS(\Omega),
\end{align}
where it is decomposed into its real, $\Gamma (\Omega)=2\text{Re}[\widehat{\mathcal{W}}(\Omega)]$, and imaginary, $S(\Omega)=\text{Im}[\widehat{\mathcal{W}}(\Omega)]$, parts respectively. Employing the symmetry $\overline{\mathcal{W}}(s)=\mathcal{W}(-s)$, where the overline denotes complex conjugation, it can be shown that
\begin{align}\label{transition:rate}
    \Gamma (\Omega):=\int_{-\infty}^{+\infty}ds\,e^{-i\Omega s}\mathcal{W}(s).
\end{align}
The above Fourier transform of the Wightman function defines the \emph{transition rate} \cite{Birrell:Davies} of the detector, i.e., the detector's probability to transition between its energy levels per unit time, due to its interaction with the field.
 
We assume that initially the detector and the field are in an uncorrelated state, $\rho_{\text{tot}}(0)=\rho(0)\otimes\ket{\Psi}\bra{\Psi}$.
In the Born-Markov approximation, the Schrödinger picture evolution of the reduced density matrix of the detector is described by the second-order master equation
\cite{DMCA,BJA:DM}:
\begin{align}\label{master:eq}
    \dot{\rho}(\tau)=-i[H_{\text{eff}},\rho(\tau)]+\mathcal{D}(\rho(\tau)),
\end{align}
where the dissipator reads
\begin{align}
    \mathcal{D}(\rho(\tau))&=\lambda^2\Gamma(-\Omega)\left(\sigma_-\rho\sigma_+-\frac{1}{2}\{\sigma_+\sigma_-,\rho\}\right)\nonumber\\&+\lambda^2\Gamma (\Omega)\left(\sigma_+\rho\sigma_--\frac{1}{2}\{\sigma_-\sigma_+,\rho\}\right),
\end{align}
and the effective Hamiltonian is $H_{\text{eff}}=\widetilde{\Omega}\sigma_z/2$, where $\widetilde{\Omega}:=\Omega+\lambda^2(S(-\Omega)-S(\Omega))$ is the Lamb shifted frequency.

For a general density matrix of the detector
\begin{align}\label{density:matrix}
    \rho(\tau)=\frac{1}{2}\left(I+\boldsymbol{r}(\tau)\cdot\boldsymbol{\sigma}\right),
\end{align}
where $\boldsymbol{r}$ is the Bloch vector,
we solve the master equation \eqref{master:eq} to obtain
\begin{align}\label{master:eq:sol}
    r_1(\tau)&=e^{-\lambda^2\frac{\gamma}{2}\tau}\bigg(r_1(0)\cosh(\widetilde{\Omega}\tau)-r_2(0)\sinh(\widetilde{\Omega}\tau)\bigg),\nonumber\\
    r_2(\tau)&=e^{-\lambda^2\frac{\gamma}{2}\tau}\bigg(r_2(0)\cosh(\widetilde{\Omega}\tau)-r_1(0)\sinh(\widetilde{\Omega}\tau)\bigg),\nonumber\\
    r_3(\tau)&=e^{-\lambda^2\gamma\tau}r_3(0)+\kappa\left(1-e^{-\lambda^2\gamma\tau}\right),
\end{align}
where we have defined  the decay rate
\begin{align}
    \gamma:=\Gamma(\Omega)+\Gamma(-\Omega),
\end{align}
and, for simplicity, the ratio
\begin{align}
     \kappa:=\frac{\Gamma(\Omega)-\Gamma(-\Omega)}{\Gamma(\Omega)+\Gamma(-\Omega)}.
\end{align}

\section{UDW detector in BTZ spacetime}\label{sec:btz}

Here, we consider a UDW detector interacting with a massless, conformally coupled scalar quantum field $\varphi(\x)$ in a (2+1)-dimensional BTZ black hole background \cite{BTZ1,BTZ2}.
The (spinless) BTZ black hole is described by the metric,
\begin{align}\label{BTZ:metric}
ds^2=-\frac{r^2-r_H^2}{\ell^2}dt^2+\frac{\ell^2}{r^2-r_H^2}dr^2+r^2d\phi^2,
\end{align}
 in Schwarzchild-like coordinates with $t\in(-\infty,\infty)$, $r\in(0,\infty)$, $\phi\in[0,2\pi)$. The metric is a vacuum solution to the Einstein field equations with a negative cosmological constant, $\Lambda=-1/\ell^2$, where $\ell$ is the anti-de Sitter length. It is asymptotic to anti-de Sitter ($\text{AdS}$) space  and admits a horizon at $r_H=\sqrt{M}\ell$, where $M$ represents the mass of the black hole.

The Wightman function $\mathcal{W}_{\text{BTZ}}(\x,\x')=\expval{0|\varphi(\x)\varphi(\x')|0}$ for a conformally coupled scalar field in the vacuum state $\ket{0}$ in the BTZ spacetime is obtained in terms of the corresponding vacuum two-point correlation function  $\mathcal{W}_{\text{AdS}}(\x,\x')$ in anti-de Sitter space by means of the method of images and is given by  \cite{Lifschytz:Ortiz,Carlip:1995}:
\begin{align}
    &\mathcal{W}_{\text{BTZ}}(\x,\x')=\sum_{n=-\infty}^{+\infty}\mathcal{W}_{\text{AdS}}(\x,Z^n\x')\\&=\frac{1}{4\sqrt{2}\pi\ell}\sum_{n=-\infty}^{+\infty}\left(\frac{1}{\sqrt{\sigma(\x,Z^n\x')}}-\frac{\zeta}{\sqrt{\sigma(\x,Z^n\x')+2}}\right)\nonumber
\end{align}
where $Z$ is the action $Z:(t,r,\phi)\mapsto(t,r,\phi+2\pi)$, the parameter $\zeta\in\{-1,0,1\}$ specifies respectively the Neumann $(\zeta=-1)$, transparent $(\zeta=0)$, and Dirichlet $(\zeta=1)$ boundary conditions imposed on the field at spatial infinity, and
\begin{widetext}
\begin{align}
\sigma(\x,Z^n\x')=\frac{rr'}{r_H^2}\cosh\left(\frac{r_H}{\ell}(\Delta\phi-2\pi n)\right)-1-\frac{\sqrt{(r^2-r_H^2)(r'^2-r_H^2)}}{r_H^2}\cosh\left(\frac{r_H}{\ell^2}\Delta t\right),
\end{align}
\end{widetext}
with $\Delta\phi=\phi-\phi'$ and $\Delta t=t-t'$.
Dirichlet and Neumann boundary conditions are imposed on the field to address the lack of global hyperbolicity in AdS spacetime (and consequently in BTZ spacetime), ensuring a well-defined quantization scheme. Alternatively, a quantization scheme can also be defined without imposing boundary conditions on AdS, known as the transparent boundary condition \cite{Isham}.

\subsection{Response of the detector}
Then, we consider that the UDW detector is kept in a fixed position outside the black hole horizon, that is, following the worldline $\mathsf{x}(\tau)\!=\!((-g_{00})^{-1/2}\tau,r,\phi)$.  The correlation function $\mathcal{W}_{\text{BTZ}}(\x,\x')$ is  shown \cite{Lifschytz:Ortiz, Carlip:1995} to be periodic in imaginary time for each of the three boundary conditions, satisfying the Kubo-Martin-Schwinger (KMS) condition \cite{Kubo,Martin:Schwinger},
\begin{align}
    \mathcal{W}_{\text{BTZ}}(\tau)=\mathcal{W}_{\text{BTZ}}(-\tau-i\beta),
\end{align}
in the exterior of the black hole, with a period $\beta$ given by
\begin{align}\label{KMS:temp}
    \beta^{-1}=T=\frac{r_H}{2\pi\ell\sqrt{r^2-r_H^2}}.
\end{align}
$T$ is interpreted as a local temperature, corresponding to the Tolman temperature \cite{tolman1934relativity}, $T=(g_{00})^{-1/2}T_0,$ with $T_0=r_H/(2\pi\ell^2)$ being the Hawking temperature of the black hole. This suggests that the Wightman function is a thermal correlation function, indicating that the conformally coupled quantum field in black hole spacetime behaves as a heat bath at temperature $T$, with the corresponding vacuum state being the Hartle-Hawking vacuum \cite{H:H}.

For the static detector outside the horizon, the transition rate is given by \cite{Lifschytz:Ortiz}:
\begin{align}\label{UDW:trans:rate}
\Gamma(\Omega)=\frac{1}{2}\left(\frac{1}{e^{\Omega/T}+1}\right)&\sum_{n=-\infty}^{+\infty}\left(P_{-\frac{1}{2}+i\frac{\Omega}{2\pi T}}(\cosh\alpha_n^-)\right.\nonumber\\&\left.-\zeta P_{-\frac{1}{2}+i\frac{\Omega}{2\pi T}}(\cosh\alpha_n^+)\right),
\end{align}
where $P_{\nu}(x)$ is the associated Legendre function of the first kind \cite{gradshteyn2014table},  and
\begin{align}
\cosh\alpha_n^{\mp}&=\frac{r_H^2}{r^2-r_H^2}\left[\frac{r^2}{r^2_h}\cosh\left(\frac{r_H}{\ell}2\pi n\right)\mp 1\right]\nonumber\\
&=(1+4\pi^2\ell^2 T^2)\cosh(2\pi n \sqrt{M})\mp 4\pi^2\ell^2 T^2.
\end{align}
The $n=0$ term is independent of the mass parameter and corresponds to the transition rate of a uniformly accelerating detector in anti-de Sitter space. The $n\neq0$ terms represent genuine black hole contributions, with their effects being most significant for small $M$ \cite{HENDERSON:2020}. Besides, the Lamb shift of the detector, resulting from its interaction with the conformal field,  has been analyzed in \cite{BTZ:Lamb}.
\begin{figure}[!b]
    \centering
    \includegraphics[width=0.95\linewidth]{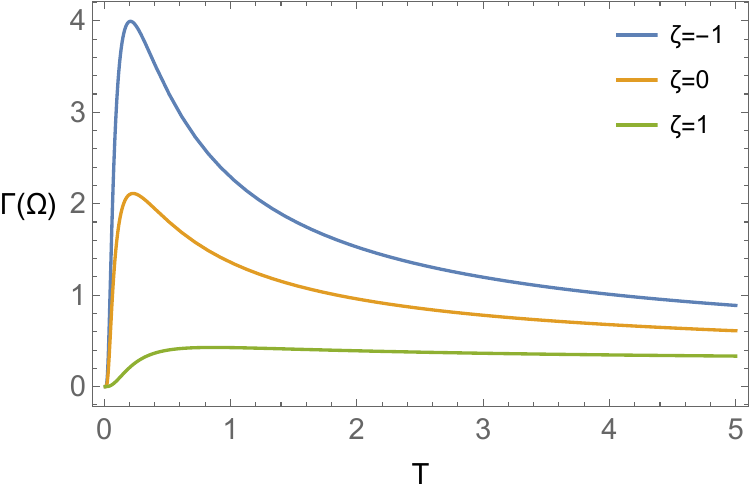}
    \caption{\justifying The transition rate, Eq. \eqref{UDW:trans:rate}, of a static UDW  detector interacting with a conformally coupled field in a (2+1)-BTZ black hole background, as a function of the KMS temperature T, for the three boundary conditions satisfied by the field at spatial infinity. Here, $M=0.01$, and $\Omega=0.1$.}
    \label{fig:transition:rate}
\end{figure}
The Legendre functions satisfy the relation $P_{-\frac{1}{2}+i\xi}(x)=P_{-\frac{1}{2}-i\xi}(x)$, from which it follows that for all boundary conditions the transition rate is thermal, i.e., it satisfies the detailed balance condition
\begin{align}
    \Gamma(\Omega)=e^{-\beta\Omega}\Gamma(-\Omega).
\end{align}
As a result, in the long-time asymptotic limit, the detector reaches a thermal equilibrium state \cite{BJA:DM},
\begin{align}\label{asympt:state}
    \rho_{\infty}=\frac{1}{2}\begin{pmatrix}
1-\tanh(\beta\Omega/2) & 0 \\
0 & 1+\tanh(\beta\Omega/2)
\end{pmatrix},
\end{align}
at the KMS temperature $T$. Hereafter, we measure temperature $T$ and energy $\Omega$ in units of $\ell^{-1}$ by setting $\ell=1$. For convenience, we also rescale the time variable as $\lambda^2\tau\to\tau$, meaning that time $\tau$ is measured in units of $\lambda^{-2}$ (the coupling strength $\lambda$ has dimensions of $\text{(energy)}^{1/2}$).
We remark that while the detector's transition rate, given by Eq. \eqref{UDW:trans:rate}, strongly depends on the type of boundary conditions--Neumann, transparent, or Dirichlet-- satisfied by the field in the BTZ spacetime (see, Fig. \ref{fig:transition:rate}, and the relevant analysis in \cite{HENDERSON:2020}), its asymptotic state remains unique, regardless of these conditions. This highlights the importance of considering the finite-time dynamics of a relativistic quantum Otto thermal machine within the open-system framework described in Sec. \ref{sec:open:dyn} as the choice of boundary conditions significantly affects the machine output power.

\section{Quantum Otto machines in BTZ spacetime}\label{BTZ:Otto}
Now we utilize the thermal properties of the conformal field in the BTZ black hole background to model the field as a heat reservoir, with the UDW detector as the working medium, and investigate the performance of a relativistic quantum Otto machine.
To investigate the thermodynamic properties of the Otto cycle, we define respectively the changes in heat and work \cite{Alicki_1979},
\begin{align}
    \expval{Q}=\int_0^\tau dt\, \text{tr}\bigg(\dot\rho(t)H(t)\bigg),
\end{align}
\begin{align}
    \expval{W}=\int_0^\tau dt\, \text{tr}\bigg(\rho(t)\dot H(t)\bigg),
\end{align}
for a quantum system with density matrix $\rho(t)$ evolving under a time-dependent Hamiltonian $H(t)$, and interacting with an external environment. Then, the first law of thermodynamics reads
\begin{align}
    \Delta U=\int_0^\tau dt\, \frac{d}{dt}\text{tr}\bigg(\rho(t)H(t)\bigg)=\expval{Q} + \expval{W},
\end{align}
where $\Delta U$ is the change in the internal energy of the system.
We start with the initial state of the working medium, a qubit detector, prepared with an energy gap $\Omega_c$ between its two energy levels as follows, 
\begin{align}
    \rho_{in}=\frac{1}{2}\begin{pmatrix}
1-\tanh(\beta_c\Omega_c/2) & 0 \\
0 & 1+\tanh(\beta_c\Omega_c/2)
\end{pmatrix}.
\end{align}
The quantum Otto thermodynamic cycle comprises four consecutive strokes, as described below:

\textit{(1.)} Adiabatic expansion -- The working medium undergoes an adiabatic expansion, during which its energy gap increases from $\Omega_c$ to $\Omega_h$, with $\Omega_h>\Omega_c$, while it
remains isolated from the field environment. The qubit remains in the initial state $\rho_{\text{in}}$ and there is no heat exchange during the process. Thus, the change in the system internal energy is identified as work, given by
\begin{align}
    \expval{W_1}= \frac{\Omega_h-\Omega_c}{2}\text{tr}\left(\rho_{in}\sigma_3\right),
\end{align}
where $\text{tr}\left(\rho_{in}\sigma_3\right)=-\tanh(\beta_c\Omega_c/2).$

\textit{(2.)} Hot isochore -- The working medium, with free Hamiltonian $H_0\!=\!\Omega_h\sigma_3/2$, is coupled to a conformally coupled quantum field in the BTZ black hole spacetime, at the inverse KMS temperature $\beta_h^{-1}$, as given by Eq. \eqref{KMS:temp}, with $\beta^{-1}_h>\beta^{-1}_c$. 
Due to the interaction with the field, the populations of the
energy levels of the system change and its density matrix evolves from $\rho_{in}$ to $\rho_h(\tau_h)$. The system dynamics is described by the second-order master equation, Eq. \eqref{master:eq}.
The mean heat exchanged with the field during the process is 
\begin{align}
    \expval{Q_h}=\frac{\Omega_h}{2}\text{tr}\left[(\rho_h(\tau_h)-\rho_{in})\sigma_3\right],
\end{align}
where
\begin{align}
   \text{tr}\left(\rho_h(\tau_h)\sigma_3\right)&= r_{3,h}(\tau_h)=\, -e^{-\gamma_h\tau_h}\tanh(\beta_c\Omega_c/2)\nonumber\\&-(1-e^{-\gamma_h\tau_h})\tanh(\beta_h\Omega_h/2).
\end{align}

\textit{(3.)} Adiabatic compression -- The working medium is isolated from the field environment and undergoes an adiabatic contraction, during which its energy gap decreases back to its initial value $\Omega_c$. The mean work done during this step is
\begin{align}
        \expval{W_3}=\frac{\Omega_c-\Omega_h}{2}\text{tr}\left(\rho_h(\tau_h)\sigma_3\right).
    \end{align}

\textit{(4.)} Cold isochore --  The working medium interacts with a conformally coupled quantum field in the BTZ background spacetime  at temperature $\beta_c^{-1}$. 
The population of the system energy levels change as a result of the interaction, and its state evolves as $\rho_c(\tau_c)$. The mean heat transferred during the process reads 
    \begin{align}
        \expval{Q_c}=\frac{\Omega_c}{2}\text{tr}\left[(\rho_c(\tau_c)-\rho_h(\tau_h))\sigma_3\right],
    \end{align}
where
\begin{align}
   \text{tr}\left(\rho_c(\tau_c)\sigma_3\right)&=r_{3,c}(\tau_c)=\, e^{-\gamma_c\tau_c}\text{tr}\left(\rho_h(\tau_h)\sigma_3\right)\nonumber\\&-(1-e^{-\gamma_c\tau_c})\tanh(\beta_c\Omega_c/2).
\end{align}

At the end of the cycle, the working medium should be back to its initial state, to obtain a closed thermodynamic cycle (that is,  $\rho_c(\tau_c)=\rho_{\text{in}}$). This is satisfied in the long-time limits, $\gamma_h\tau_h\gg 1,$ and $\gamma_c\tau_c\gg 1$, during steps (2) and (4) of the cycle, when the detector reaches its asymptotic equilibrium state. In this asymptotic regime, the heat and work changes read
\begin{align}\label{work:heat:asympt}
    \expval{Q_h}&=\frac{\Omega_h}{2}\bigg(\tanh(\beta_c\Omega_c/2)-\tanh(\beta_h\Omega_h/2)\bigg),\nonumber\\
        \expval{W_3}&=\frac{\Omega_h-\Omega_c}{2}\tanh(\beta_h\Omega_h/2),\nonumber\\
    \expval{Q_c}&=\frac{\Omega_c}{2}\bigg(\tanh(\beta_h\Omega_h/2)-\tanh(\beta_c\Omega_c/2)\bigg).
\end{align}

\subsection{Quantum Otto heat engine}
In this case, heat is absorbed from the hot environment, $\expval{Q_h}\geq 0$, and flows into the cold $\expval{Q_c}\leq 0$. The work is extracted from the working medium for positive total work $W\!=\!-(\expval{W_1}+\expval{W_3})>0$. This implies the condition $\Omega_h/\Omega_c<T_h/T_c$. 
The efficiency $\eta$ of a quantum Otto heat engine is defined as the ratio of the total work output to the absorbed heat, then reads
\begin{align}
    \eta=-\frac{\expval{W_1}+\expval{W_3}}{\expval{Q_h}}=1-\frac{\Omega_c}{\Omega_h},
\end{align}
and the power output as the total work to the duration $\tau_{\text{cycle}}$ of the cycle,
\begin{align}\label{power:out}
    P=-\frac{\expval{W}_{\text{tot}}}{\tau_{\text{cycle}}}.
\end{align}
Without loss of generality, we explicitly account for the durations of the heating and cooling strokes, writing $\tau_{\text{cycle}}\!=\!\tau_h+\tau_c$, while neglecting the duration of the adiabatic strokes \cite{Deffner:EMP}.

\begin{figure}[t!]
    \centering
 \includegraphics[width=0.95\linewidth]{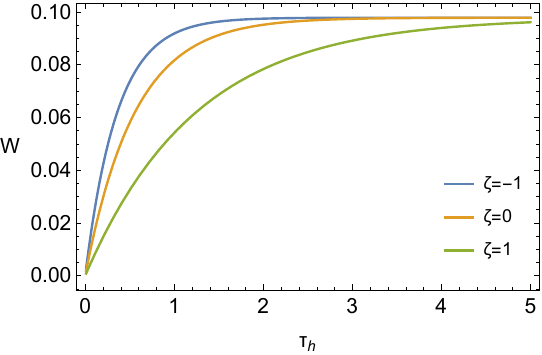}
    \caption{\justifying Total work output of the heat engine as a function of the evolution time $\tau_h$, for the Neumann ($\zeta=-1$), transparent $(\zeta=0)$ and Dirichlet $(\zeta=1)$ boundary conditions, satisfied by the conformally coupled field at spatial infinity. The parameters used for the plot are $M=0.01$,  $\Omega_h=1$, $\Omega_c=0.1$, $T_h=2$, $T_c=0.1$.}
    \label{work:tot}
\end{figure}

In Fig. \ref{work:tot},  we plot the total work done by the qubit detector as a function of the evolution time $\tau_h$ for the three different boundary conditions satisfied by the conformally coupled scalar field in the BTZ spacetime. We observe that the total work reaches its maximum output in the long-time limit, asymptotically approaching the values given in Eqs. \eqref{work:heat:asympt}, which are independent of the boundary conditions.  However, for finite engine operation times $\tau_h$, the work output strongly depends on the boundary conditions. At earlier times, the Neumann boundary condition ($\zeta=-1$)  yields the highest work output. This case also surpasses the other two boundary conditions in reaching the maximum asymptotic value in earlier times.


The maximum efficiency of a heat engine is determined by the Carnot efficiency \cite{Callen}, given by $\eta_C=(T_h-T_c)/T_h$. However, reaching Carnot efficiency requires an infinite slow process to ensure perfect adiabatic evolution and complete thermalization with the heat reservoirs during each of the four strokes of the cycle, resulting in zero power output. A more practical measure for analyzing the performance of a heat engine is the \emph{efficiency at maximum power} (EMP), which is determined by first optimizing the power output with respect to a system parameter and then calculating the efficiency at that optimized power. Curzon and Ahlborn \cite{Curzon:Ahlborn} derived the efficiency of a Carnot engine operating at maximum power as $\eta_{CA}=1-\sqrt{T_c/T_h}$. It balances the trade-off between high power output and low entropy production, incorporating the environmental impact of heat engines.

The \emph{ecological criterion}, introduced by Angulo-Brown \cite{Angulo:Brown}, is an alternative optimization criterion for evaluating the performance of heat engines.  It embodies a compromise between maximizing useful work and minimizing the environmental cost of waste heat and entropy production.  This criterion is implemented using the ecological function:
\begin{align}\label{ecological:eng}
    E=P-T_cS,
\end{align}
where $S$ is total entropy change per unit time of the cycle, given by $S=\beta_c\expval{Q_c}-\beta_h\expval{Q_h}$.

\begin{figure}[t!]
    \centering    \includegraphics[width=0.95\linewidth]{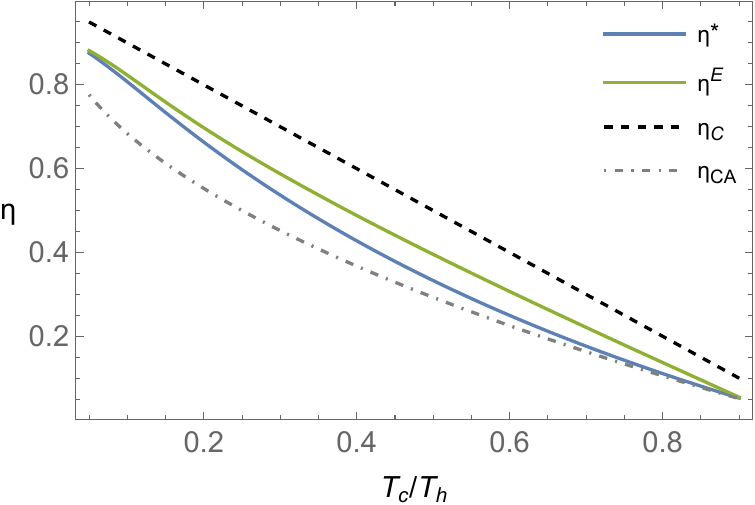}
    \caption{\justifying The efficiency of the relativistic Otto heat engine as a function of the temperature ratio $T_c/T_h$. Here, $\eta^*$ is the engine's efficiency at maximum power, while $\eta^E$ denotes the efficiency at maximum ecological function, both optimized with respect to the qubit's energy gap $\Omega_h$. Additionally, $\eta_C$ is Carnot efficiency, and $\eta_{CA}$ the Curzon-Ahlborn efficiency. The parameters used for the plots are: $M\!=\!0.01$, $\zeta\!=\!-1$, $\Omega_c\!=\!0.1$, $\tau_h\!=\!0.2$, $\tau_c\!=\!0.5$.}
    \label{fig:heat:eng:opt}
\end{figure}
We analyze the performance of the relativistic Otto heat engine, optimizing both the power output, Eq. \eqref{power:out}, and the ecological function, Eq. \eqref{ecological:eng}, with respect to the energy gap $\Omega_h$ of the qubit detector. The optimization is performed for fixed values of the parameters $M$, $\zeta$, $T_h$, $\Omega_c$, $\tau_h$ and $\tau_c$, by numerically solving  equations $\partial P/\partial\Omega_h=0$ and $\partial E/\partial\Omega_h=0$. We then evaluate the efficiency at maximum power $\eta^*$, and at maximum ecological function $\eta^E$. In Fig. \ref{fig:heat:eng:opt}, we present the performance of the heat engine as a function of the temperature ratio $T_c/T_h$.

We verify numerically that the efficiency at maximum power coincides for the three different boundary conditions. As a result, we conclude that the Neumann boundary condition represents the optimal case, as it yields the highest power output for finite-time operation of the heat engine. The EMP also exceeds the Curzon-Ahlborn $\eta_{CA}$ in the intermediate temperature ratio range, which is similar to the result of the BTZ black hole Brayton engine cycle \cite{Ferketic:Deffner}. In addition, we observe that in the intermediate temperature ratio range the efficiency at the maximum ecological function $\eta^E$ is different to the EMP $\eta^*$. When the ecological function is maximum, the heat engine releases a small amount of heat to the surrounding environment, making its operation  more ecological. We remark that both the finite-time EMP values $\eta^E$ and $\eta^*$ closely resemble the ones analytically obtained in \cite{singh:OA} for a qubit heat engine in the high-temperature regime, and converge to the same values in the long-time asymptotic limit. 


\subsection{Quantum Otto refrigerator}
The Otto thermodynamic cycle operates as a refrigerator when heat is absorbed from the cold environment, $\expval{Q_c}\geq 0$, and flows into the hot environment, $\expval{Q_h}\leq 0$. That is, satisfying the condition $\Omega_h/\Omega_c> T_h/T_c$.

\begin{figure}[t!]
    \centering
 \subfloat[$\tau_h\!=\!0.3$]{\includegraphics[width=0.95\linewidth]{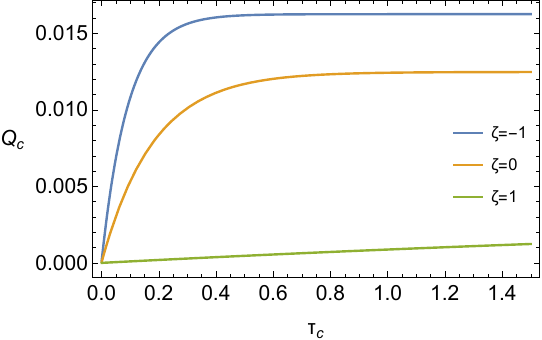}}\\
 \subfloat[$\tau_h\!=\!10$]{\includegraphics[width=0.95\linewidth]{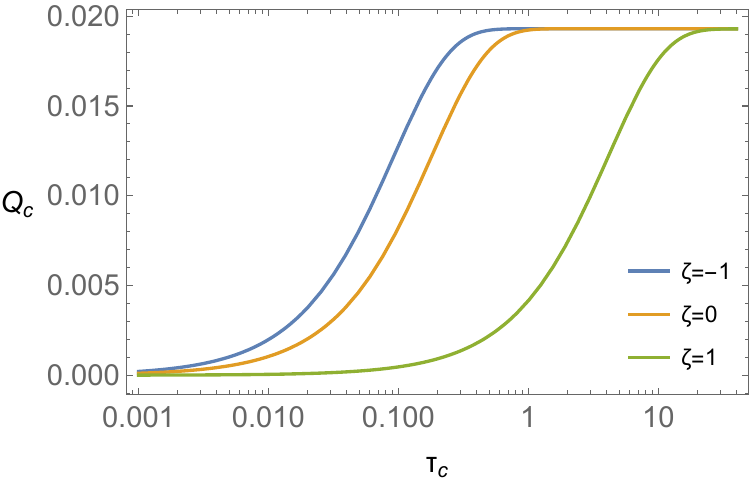}\label{fig:heat:b}}
    \caption{\justifying Cooling power of the relativistic Otto refrigerator as a function of the evolution time $\tau_c$  for the Neumann $\zeta=-1$, transparent $(\zeta=0)$ and Dirichlet $\zeta=1$ boundary conditions, when (a) $\tau_h=0.3$ and (b) $\tau_h=10$. The other parameters used for the plots are: $M=0.01$,  $\Omega_h\!=\!0.5$, $\Omega_c\!=\!0.1$, $T_h\!=\!0.2$, $T_c\!=\!0.1$.}
    \label{fig:heat}
\end{figure}

In Fig. \ref{fig:heat}, we plot the cooling power $\expval{Q_c}$, as a function of the evolution time $\tau_c$ for Newmann, transparent and Dirichlet boundary conditions, considering (a) a finite interaction time $\tau_h$, and (b) the long-time limit during the hot isochore stroke. In Fig. \ref{fig:heat:b}, we observe that the absorbed heat reaches its maximum value in the long-time limits, asymptotically approaching the value given in Eq. \eqref{work:heat:asympt}. For the Dirichlet boundary condition, achieving this asymptotic value requires  longer times $\tau_c$ than the other two cases. This results from the lower transition ratesbetween the detector's energy levels in the low-temperature regime, as shown in Fig. (\ref{fig:transition:rate}). 
Similarly to the heat engine scenario, for finite-time operation of the refrigerator, the Neumann boundary condition again proves to be the optimal choice,  as it yields the highest cooling power. It also reaches its maximum asymptotic value much faster than the Dirichlet boundary condition.

The \emph{coefficient of performance} (COP) $\varepsilon$ of an Otto refrigerator is defined as the ratio of heat absorbed from the cold environment to the total amount of work:
\begin{align}
    \varepsilon=\frac{ \expval{Q_c}}{\expval{W_1}+\expval{W_3}}=\frac{\Omega_c}{\Omega_h-\Omega_c},
\end{align}
where the second equality holds for a closed thermodynamic cycle. The maximum COP of a refrigerator is $\varepsilon_C=T_c/(T_h-T_c)$ \cite{Callen}. 

To optimize the performance of a refrigerator,  we employ the \emph{figure of merit} $\chi$ \cite{Abah_2016}, defined as the product of the COP and the cooling power $\expval{Q_c}$ to the duration $\tau_{cycle}$ of the cycle: 
\begin{equation}\label{figure:merit}
\chi=\frac{\varepsilon\expval{Q_c}}{\tau_{cycle}}.
\end{equation} 
On the other hand, the ecological criterion for refrigerators is implemented using the ecological function \cite{singh:OA,Fernandez_2022}
\begin{align}\label{ecolog:refrig}
    E'=Q_c-\varepsilon_C T_h S.
\end{align}

 
We analyze the performance of the relativistic Otto refrigerator by optimizing the figure of merit, Eq. \eqref{figure:merit}, and the ecological function, Eq. \eqref{ecolog:refrig}, with respect to the energy gap $\Omega_c$ of the qubit detector. The optimization is performed for fixed values of the parameters $M$, $\zeta$, $T_h$, $\Omega_h$, $\tau_h$ and $\tau_c$, by numerically solving equations $\partial \chi/\partial\Omega_c=0$ and $\partial E'/\partial\Omega_c=0$.  We now evaluate the COP at the maximum $\chi$ figure of merit, $\varepsilon^*$, and the maximum ecological function, $\varepsilon^E$. 

\begin{figure}[t!]
    \centering
    \includegraphics[width=0.95\linewidth]{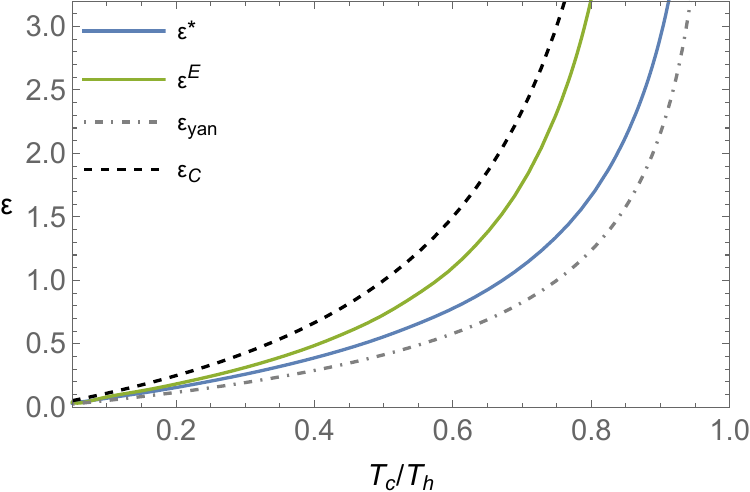}
    \caption{\justifying Coefficient of performance (COP) of the relativistic Otto refrigerator as a function of the temperature ratio $T_c/T_h$ (case $\zeta=-1$). Here, $\varepsilon^*$ is the COP at maximum figure of merit and $\varepsilon^E$ the COP at maximum ecological function, both optimized with respect to the qubit's energy gap $\Omega_c$. Also, $\varepsilon_C$ is the Carnot and  $\varepsilon_{yan}$ the Yan-Chen COP, respectively. The parameters used for the plot are: $M=0.01$, $\Omega_h=0.5$, $\tau_h=\tau_c=0.2$.}
    \label{fig:energy:opt:refr}
\end{figure}

In Fig. \ref{fig:energy:opt:refr}, we present the COP of the refrigerator as a function of the temperature ratio $T_c/T_h$.
We verify that the COP at maximum figure of merit coincides for the three different boundary conditions satisfied by the quantum field. Thus, we again conclude that the Neumann boundary condition represents the optimal case, and thus absorbs maximum heat from the cold field bath for finite-time operations. We observe that the COP at maximum ecological function $\varepsilon^E$ is higher than the COP at maximum figure of merit $\varepsilon^*$, with both exceeding the Yan-Chen COP \cite{Yan:Chen}, which is the counterpart of the Curzon-Ahlborn efficiency for refrigerators, and is given by $\epsilon_{\text{yan}}=\sqrt{1+\epsilon_C}-1$. We note that both the finite-time COP values $\varepsilon^*$ and $\varepsilon^E$ closely resemble the ones that are analytically obtained in \cite{singh:OA} for a qubit refrigerator in the high-temperature regime, and coincide in the long-time asymptotic limit. 

\section{Concluding remarks}\label{conclusions}

We presented a general framework for studying the finite-time operation of relativistic quantum thermal machines, with a focus on their energy optimization. We applied this framework to study a UDW detector coupled to a quantum field in the Hartle-Hawking state in a BTZ black hole background spacetime, implementing an Otto thermodynamic cycle with the detector as the working medium. By modeling the UDW detector as an open quantum system, with the field acting as a thermal bath, we described its finite-time dynamics in each stroke of the cycle. This approach allowed us to examine how the performance of an Otto heat engine or refrigerator is explicitly affected by Neumann, transparent, and Dirichlet boundary conditions imposed on the field. We identified the Neumann boundary condition as the optimal choice, yielding the highest work output and cooling power.

Since achieving Carnot efficiency represents an idealized case requiring perfect adiabatic evolution and complete thermalization at each stage of the thermodynamic cycle, we focused on optimizing the Otto thermal machine’s performance in terms of power and ecological impact. This approach can be directly applied to detectors following different stationary trajectories in various curved spacetime backgrounds. Thus, our work provides a comprehensive analysis of quantum thermal machines and their energy optimization in relativistic settings.

 Finally, we note that several studies have proposed the realization of black holes in analogue experiments in laboratory settings, employing fluid dynamics, optics, or Bose-Einstein condensates to simulate event horizons and the emitting Hawking radiation (see, e,g., \cite{Uwe1,Uwe2,Ulf, de_Nova_2019,almeida2023analogue} and references therein). In particular, analogue models of (2+1) BTZ black holes have been proposed in curved graphene sheets \cite{kandemir2020hairy} and photon fluids \cite{chen2023optical,chen2}. This suggests the possibility of exploring thermal machines in black hole analogue setups.


\section{Acknowledgments}

This work was supported by the UK Research and Innovation Engineering and Physical Sciences Research Council [grant number EP/Z002796/1].


\bibliography{references}

\end{document}